\documentclass{iaus}
\usepackage{graphicx}

\title[Halo missing baryons] 
{The halo by halo missing baryon problem}

\author[McGaugh]   
{Stacy S. McGaugh
}
  
\affiliation{Department of Astronomy,
  University of Maryland, College Park, MD 20742, USA
 \break email: ssm@astro.umd.edu}

\pubyear{2007}
\volume{244}  
\pagerange{119--126}
\date{?? and in revised form ??}
\jname{Dark Galaxies \& Missing Baryons}
\editors{J. Davies, ed.}
\begin{document}

\maketitle

\begin{abstract}
The global missing baryon problem --- that the sum of observed
baryons falls short of the number expected form BBN --- is well known.
In addition to this, there is also a local missing baryon problem that
applies to individual dark matter halos.  This halo by halo missing 
baryon problem is such that the observed mass fraction of baryons 
in individual galaxies falls short of the cosmic baryon fraction.
This deficit is a strong function of circular velocity.  I give an empirical
estimate of this function, and note the presence of a critical scale
of $\sim 900\;\textrm{km}\,\textrm{s}^{-1}$ therein.  I also briefly review 
$\Omega_b$ from BBN, highlighting the persistent tension between 
lithium and the CMB, and discuss some pros and cons of individual
galaxies and clusters of galaxies as potential reservoirs of dark baryons.
\end{abstract}

\firstsection 
\section{Introduction}

Modern cosmology suffers a hierarchy of missing mass problems.  
Most famously, there appears to be more gravitating mass than can
be accounted for with baryons while maintaining consistency with
Big Bang Nucleosynthesis (BBN):
\begin{equation}
\Omega_b < \Omega_m.
\end{equation}
This, together with the need to grow large scale structure, leads to the
inference of non-baryonic cold dark matter (CDM).  

In addition to the dynamical missing mass problem, there is also the 
missing baryon problem (e.g., \cite{PS, FHP}).  
Our inventory of known baryons in the local universe, summing over
all observed stars, gas, etc., comes up short of the total anticipated
from BBN:
\begin{equation}
\Omega_* + \Omega_g + \dots < \Omega_b.
\end{equation}
For example, \cite{Bell} estimate that the sum of stars and cold
gas is only $\sim 3\%$ of $\Omega_b$.
While there now seems to be a good chance that many of the missing
baryons are in the form of highly ionized gas in the warm--hot 
intergalactic medium (the WHIM --- see Mathur, these proceedings),
we are still far from being able to give a confident accounting of where
all the baryons reside.  Indeed, there could be multiple distinct reservoirs 
in addition to the WHIM, each comparable to the mass in stars, within
the current uncertainties.

Here I highlight the halo by halo missing baryon problem.
In addition to the global missing baryon problem, there seems to be a
mismatch between the mass of baryons in individual galaxies 
and the mass of their host halos.
For the amount of dark matter we infer dynamically, the mass in detected 
baryons falls short of the cosmic baryon fraction:
\begin{equation}
m_d < f_b.
\end{equation}
Here 
\begin{equation}
m_d = \frac{M_b}{M_{vir}}
\end{equation}
is the mass fraction of detected baryons in a given object,
and the cosmic baryon fraction is $f_b = 0.17$ (\cite{WMAP3}).
Unfortunately, the virial mass of any given galaxy is a rather notional
quantity, and there are many possible mappings between the observed
baryonic content and the inferred dark mass.  Nevertheless, there is
a pronounced regularity to the data that does not tumble naturally out
of our $\Lambda$CDM theory. 

\section{Big Bang Nucleosynthesis}

Before proceeding with a discussion of the missing baryon problem in
individual dark matter halos, it is worth reviewing $\Omega_b$.
It is commonly stated that we have entered the era of precision
cosmology.  If this is true, then there is a tension between 
$\Omega_b = 0.03$ and 0.04 that should concern us.

\begin{figure}
 \includegraphics[width=5in]{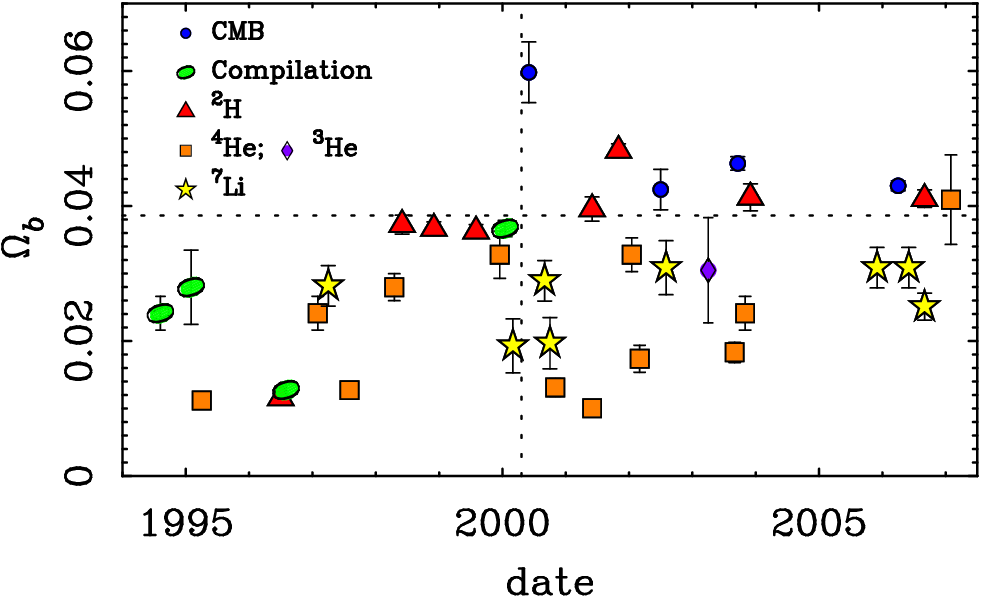}
  \caption{The baryon density from various measurements over the past
  decade, as tabulated by \cite[McGaugh (2004)]{mycmb}, with a few
  recent updates (\cite{CC, asplund, boni, newD, PLP, WMAP3}).  
  Constraints from each
  independent method are noted by different symbols.  BBN was well
  established long before the start date of this graph; previous work 
  is represented by early compilations.  This plot assumes 
  $H_0 = 72\;\textrm{km}\,\textrm{s}^{-1}\,\textrm{Mpc}^{-1}$;
  the first point is the famous $\Omega_b h^2 =0.0125$ value of
  \cite[Walker \etal\ (1991)]{oldbbn}.  
  Note that no measurement of any isotope ever
  suggested a value $\Omega_b h^2 > 0.02$ (horizontal dotted line) 
  prior to the appearance of relevant CMB data (vertical dotted line).
  With the exception of the case of no CDM 
  (\cite[McGaugh 1999, 2004]{mycmb}),
  no CMB fit has ever suggested $\Omega_b h^2 < 0.02$. 
  Measured values of some isotopes seem to have drifted upwards 
  towards the CMB values since 2000, a worrisome trend given the 
  historical tendency for cosmological measurements to track one another.}
  \label{fig:BBN}
\end{figure} 

One of the great successes of cosmology is BBN.  
The abundance of the light elements depends on a single parameter,
the baryon-to-photon ratio, that maps directly to the baryon density.
Independent measurements of $^2$H, $^4$He, and $^7$Li all indicate
nearly the same value for $\Omega_b$.  In addition, modern 
observations of the cosmic microwave background (CMB) provide
another similar constraint on the baryon density.

The consistency of the independent BBN constraints is impressive,
and I have little doubt that the basic picture is correct.  However, the
agreement is not perfect.  There seems to be a modest dichotomy 
between each of the independent methods (Fig.~\ref{fig:BBN}), 
with each preferring a slightly different value of $\Omega_b$.  
Table~\ref{tab:bbn} shows a comparison of baryon densities
by method.  Values are the median for each case from the compilation
of \cite{mycmb} assuming 
$H_0 = 72\;\textrm{km}\,\textrm{s}^{-1}\,\textrm{Mpc}^{-1}$.
The uncertainty is taken from the variance of different published 
measurements.  This tends to overstate the error claimed by individual
determinations.  For this uncertainty, the light elements are broadly
consistent.  If we look only at the latest WMAP result 
(\cite{WMAP3}), its tiny uncertainty makes
it difficult to reconcile with the helium and lithium.  I briefly review 
each method below. 
\begin{table}\def~{\hphantom{0}}
  \begin{center}
  \caption{$\Omega_b$ by method$^{*}$.}
  \label{tab:bbn}
  \begin{tabular}{lccccccc}\hline
       & W91$^{\dagger}$ & $^2$H   & $^4$He &   $^7$Li & CMB
       & W3$^{\ddagger}$  & M04$^{\star}$ \\\hline
     $\Omega_b$ & 0.024 & 0.038 & 0.021 & 0.028 & 0.046 
      & 0.0430 & 0.033 \\
     $\sigma$ & 0.005 & 0.005 & 0.008 & 0.005 & 0.007
      & 0.0014 & 0.006 \\\hline
  \end{tabular} \\
   $^*$Assumes 
  $H_0 = 72\;\textrm{km}\,\textrm{s}^{-1}\,\textrm{Mpc}^{-1}$. \\
  $^{\dagger}$Compilation of \cite[Walker \etal\ (1991)]{oldbbn}.  \\
  $^{\ddagger}$WMAP-only $\Lambda$CDM fit from $\Lambda$ website. \\
  $^{\star}$CMB without CDM --- see \cite[McGaugh (2004)]{mycmb}. \\
 \end{center}
\end{table}

\textbf{Helium:}
Measured in very low metallicity extragalactic HII regions
(e.g., \cite[Kuzio de Na\-ray \etal\ 2004]{KdN}), helium gives 
the lowest value of $\Omega_b$.  Though consistently in the vicinity of
$Y_p = 0.24$, the weak dependence of $Y_p$ on $\Omega_b$
makes it difficult to obtain a precise estimate from helium.  Moreover,
independent determinations vary by more than they should for the stated
errors.  This is likely due to the many subtle systematic effects that
afflict helium abundance determinations at the $\sim 1\%$ level
(\cite{OS}).  Thus, while the broad consistency of helium with other
determinations is comforting, it seems unlikely that it can provide a 
strong constraint on the precise value of $\Omega_b$.  We should not
take this as a free pass to simply ignore helium; \textit{all} 
determinations of $Y_p$ up to the date of my review gave
$\Omega_b < 0.035$, and I found it increasingly difficult to reconcile 
any of the helium data with $\Omega_b > 0.04$ (but see \cite{PLP}).

\textbf{Lithium:} Measured in low metallicity stars (e.g., \cite{ryan}),
lithium gives the next lowest value for $\Omega_b$.  
Unlike helium, the $^7$Li
abundance seems to be both well measured and consistent between
independent determinations.  The Spite plateau is sharply defined by
dozens of low metallicity stars.  At $\Omega_b = 0.028$, 
it is marginally consistent with deuterium but not consistent with
the CMB.  The general presumption seems therefore to be that lithium
needs to be `fixed' (e.g., \cite{Korn, Piau}), but there is nothing broken 
about it if the CMB value is overestimated or its uncertainty underestimated.

\textbf{Deuterium:} 
Of the light elements, deuterium is the most sensitive to $\Omega_b$.
Its measurement in Lyman absorption clouds in the spectra of 
high redshift quasars provides one of the most accurate
constraints on BBN.  Indeed, it came in at surprisingly low D/H, giving 
high $\Omega_b h^2 = 0.019 \pm 0.0012$ (\cite{tytler}) 
relative to long-standing results
(e.g., $\Omega_b h^2 = 0.0125 \pm 0.0025$: \cite{oldbbn}).
Unfortunately, there exist only 6 accurate, published D/H values
(fewer systems than there are points in Fig.~\ref{fig:BBN}),
one of which is discrepantly high.

\begin{figure}
 \includegraphics[width=5in]{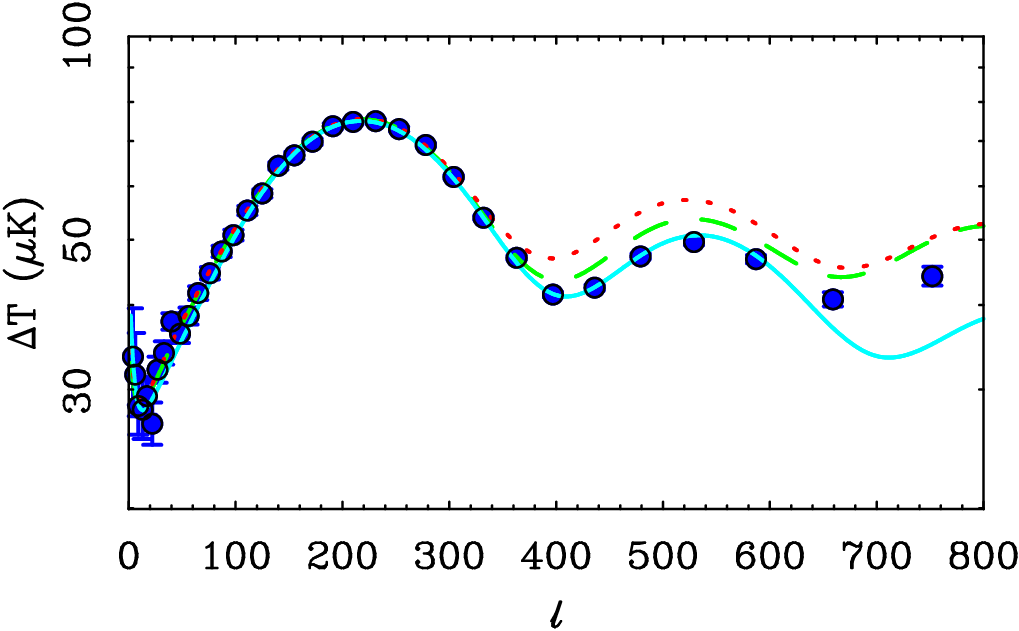}
  \caption{The effects of the baryon density and baryon fraction 
  on the CMB.  The points are the third year WMAP data 
  (\cite{WMAP3}).
  The various lines are models with $\Omega_b$ and $f_b$ set to fixed
  values (see text).  The dotted (topmost) line is set to the lithium
  value ($\Omega_b = 0.028$) with $f_b = 0.12$ (case 1).  The dashed 
  (middle) line is set to the deuterium value ($\Omega_b = 0.038$) 
  with $f_b = 0.17$ (case 2).  Case 3 is the solid (bottom) line with 
  lithium $\Omega_b$ and no CDM 
  (\cite[McGaugh 1999, 2004]{origcmb,mycmb}).  
  The CMB is exquisitely sensitive to both the absolute baryon density 
  and the baryon fraction, but also suffers degeneracies between these 
  and other parameters (especially the tilt).}
  \label{fig:CMB}
\end{figure} 

\textbf{CMB:}
The microwave background provides another constraint on
$\Omega_b$ that is independent of abundance measurements.  
It gives a value for $\Omega_b$ that is close to but persistently 
higher than all of the light elements.  The power spectrum of the CMB is
very sensitive to both the absolute density of baryons ($\Omega_b$) 
and the global baryon fraction ($f_b$).  Fig.~\ref{fig:CMB} illustrates
this sensitivity with three cases:  
(1) a plausibly low baryon density, low baryon fraction universe; 
(2) a high baryon density, high baryon fraction universe, and 
(3) a low density, no-CDM universe.

\textbf{Case (1)}
is motivated by the lithium baryon density (Table~\ref{tab:bbn}).
For this value of $\Omega_b$, $f_b = 0.12$ as observed in clusters
(\cite{clustfb}; see also Holder, these proceedings)
if $\Omega_m = 0.23$, consistent with peculiar velocity 
measurements ($\Omega_m = 0.22 \pm 0.02$: \cite{MT}).  
These three independent lines of evidence
provide a self consistent and reasonable set of cosmological constraints.
However, they badly overestimate the amplitude of the second acoustic
peak.  This problem is generic to all concordance $\Lambda$CDM models
prior to the first measurement of the second peak in 2000.

\textbf{Case (2)} 
utilizes the deuterium abundance, which gives the highest
$\Omega_b$ of all the BBN elements and the one most consistent with
fits to the CMB.  A baryon fraction $f_b = 0.17$ is adopted, 
giving a plausible $\Omega_m = 0.26$.  Though
rigged to be as consistent as possible with CMB fits without simply
adopting them, this case still over-predicts the amplitude of the second
peak, albeit by a lesser amount than case (1).

\textbf{Case (3)} 
uses the lithium $\Omega_b$ with zero CDM.
Consideration of this model
is motivated by MOND, but uses entirely conventional physics.  It provides
quite a good fit to the second peak.  Indeed, this was the only successful
\textit{a proiri} prediction of its amplitude (\cite{origcmb}).
Here it merely illustrates the sensitivity to $f_b$ as well as $\Omega_b$.

Like all models devoid of CDM, case (3) under-predicts the
amplitude of the third peak, just as un-tilted $\Lambda$CDM models
over-predict it (e.g., cases 1 and 2).  
This `low' third peak leads to the inference of a tilt 
($n \approx 0.95$), which in turn impacts the inferred baryon density,
somewhat reducing it relative to previous CMB determinations.  
In the context of MOND, the `high' third peak implies a driving term
analogous to that provided by CDM.  This occurs naturally in TeVeS-like
relativistic MOND theories (\cite{skord}), but it is unclear how this
might affect $\Omega_b$.

\subsection{BBN summary}

BBN is undoubtedly one of the shining successes of modern cosmology.  
There is broad agreement between multiple independent lines of evidence
to an accuracy rarely witnessed in cosmology.  
I have little doubt that the basic picture is correct.

In detail, there are significant differences between independent methods.
It is common to point to the great consistency between
$\Omega_b$ from BBN and the CMB as support for the
$\Lambda$CDM picture.  However, this is rather sloppy,
being only really true for deuterium.  Lithium and helium prefer noticeably 
lower values for the baryon density.  
%
%

The light elements themselves provide a self-consistent picture if we
adopt the uncertainties listed in Table~\ref{tab:bbn}.  It is the tiny
uncertainty in the CMB determination that is problematic.  The CMB
is indeed that sensitive (Fig.~\ref{fig:CMB}), but does suffer from
parameter degeneracies, in particular the tilt and the baryon fraction.
The tension between the light elements and the CMB disappears if
there is no CDM.

It appears that the true baryon density is not yet as precisely
known as might be hoped.  Nevertheless,
the missing baryon problem remains an issue at low redshift.
Whether the true baryon density is 3\% of critical or 4\% depends
on whether one gives more weight to light element 
measurements or CMB fits.  Either way, $\Omega_b$ 
exceeds the well-determined inventory of local baryons (\cite{FHP}).

\section{The baryon content of individual dark matter halos}

As well as trying to account for all the baryons in the universe,
and the integrated amount in galaxies, clusters, and the IGM 
(see Frenk; Moore, these proceedings),
it would be nice to have an accounting in individual systems.
Each dark matter halo can, to a first approximation, be thought of
as a microcosm of the whole.  As such, one would naively expect 
each halo to have the same baryon fraction as the whole universe.
On the scale of clusters of galaxies, this is approximately true.
For individual galaxies, observations depart from this ideal in
a way which we have yet to understand.

\subsection{The Milky Way}

Our own galaxy provides an illustration of the general problem.
We have a rather complete inventory of local baryons that appears
to be fairly consistent between independent estimates.  Yet when
we compare the baryonic mass to the total dynamical mass, it falls
well short of the universal baryon fraction.

For illustration, I adopt estimates of the baryonic mass
from the recent work of \cite{flynn}: 
$\textrm{M}_*^{MW} \approx 5 \times 10^{10}\;\textrm{M}_{\odot}$
and $\textrm{M}_g^{MW} \approx 10^{10}\;\textrm{M}_{\odot}$.
Most of the stellar mass is in the disk, though the spit between
disk and bulge depends somewhat on details like the disk scale
length.  The gas mass is split between molecular and atomic 
components.  Other baryonic components appear to be
negligible.  The total baryonic mass of the Milky Way is thus
$\textrm{M}_b^{MW} \approx 6 \times 10^{10}\;\textrm{M}_{\odot}$.

The baryonic masses of satellite galaxies within the Milky Way halo
add little to this sum.  Their total is probably less than the uncertainty in
the Milky Way mass,
and certainly less than $10^{10}\;\textrm{M}_{\odot}$.  It is thus
hard to imagine that the baryonic mass associated with the Milky Way
halo is any larger than $10^{11}\;\textrm{M}_{\odot}$.

The total mass of the Milky Way has been studied many times.
The answer persistently comes back in the vicinity of 
$\textrm{M}_{tot}^{MW} \approx 2 \times 10^{12}\;\textrm{M}_{\odot}$.  
There are many uncertainties in this estimate, the dominant one being
whether the most distant tracer, Leo I, is considered to be bound to the
Milky Way or not. \cite{SCB} give 
$2.5 \times 10^{12}\;\textrm{M}_{\odot}$ with Leo I and
$1.8 \times 10^{12}\;\textrm{M}_{\odot}$ without it.  
A practical lower limit is $\sim 10^{12}\;\textrm{M}_{\odot}$;
half this is already encompassed within the radius of the LMC.

Taking these numbers at face value,
\begin{equation}
m_d^{MW} = \frac{6 \times 10^{10}\;\textrm{M}_{\odot}}
{2 \times 10^{12}\;\textrm{M}_{\odot}} = 0.03.
\label{eqn:MW}
\end{equation}
This falls well short of the universal baryon fraction, $f_b = 0.17$.
Pushing the numbers to their extremes, we can place the limit
$m_d^{MW} < 0.10$.  This is at least within hailing distance of
$f_b  = 0.12$ from clusters, but still well short of the CMB value.

This raises the question:  where are all the baryons that should be
associated with the Milky Way halo?  Perhaps they have been expelled,
though the Milky Way is generally thought to be too massive to have
suffered much of this.  Another possibility is that they lurk in the halo
in some as yet undetected form, like a local WHIM.  While the WHIM
is a great candidate for many of the globally missing baryons
(see Mathur, these proceedings), it appears to fall well short of
what is needed to explain this deficit in the Milky Way (\cite{PM}),
providing perhaps $\sim 10^{10}\;\textrm{M}_{\odot}$ (\cite{CSG}).

\begin{figure}
 \includegraphics[width=5in]{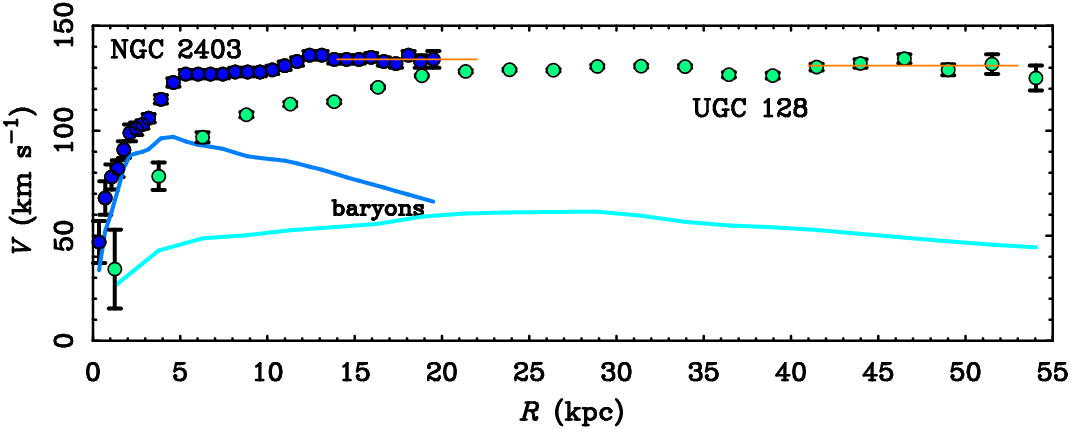}
  \caption{Rotation curves for two galaxies illustrating the circular
  velocity measured from the outer regions of extended 21 cm data
  (horizontal lines).  The distribution of the baryonic component is 
  also illustrated.  For the purposes of this review, 
  we are only concerned with the total mass in baryons
  and the characteristic circular velocity of the system.}
  \label{fig:TFpair}
\end{figure} 

The Milky Way is not unique in having $m_d < f_b$.  Similar discrepancies
are observed in many systems (\cite{MdB98a}).  Indeed, $m_d$
appears to deviate systematically from $f_b$ by an amount which
becomes increasingly severe in systems of decreasing circular velocity.

\subsection{The Baryonic Tully-Fisher Relation}

A basic tenet of CDM theory is a relation between halo mass and
circular velocity at the virial radius (\cite{SN}) of the form
\begin{equation}
M_{vir} \propto V_{vir}^3
\label{eqn:CDM}
\end{equation}
with the normalization depending on the cosmology
(\cite{MoMao}).  We can measure neither of these quantities.  
What we can measure is the baryonic mass of a given galaxy
and a characteristic circular velocity (Fig.~\ref{fig:TFpair}).

\begin{figure}
 \includegraphics[width=5in]{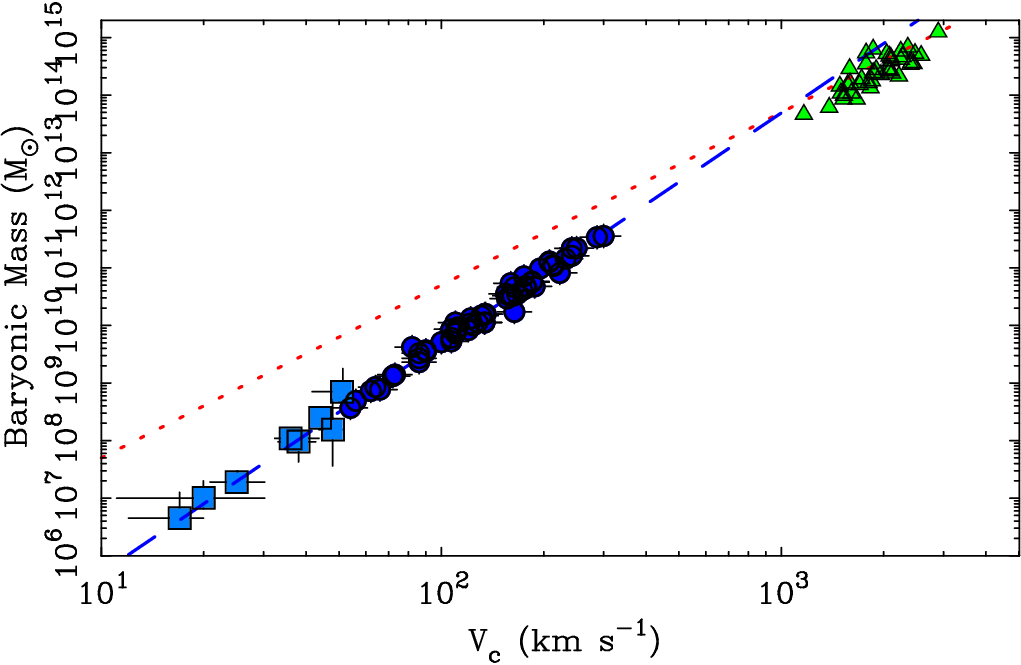}
  \caption{The baryonic mass-circular velocity relation, spanning
  the range from the tiniest dwarfs (squares) through spiral galaxies
  (circles; \cite[McGaugh 2005]{btfgood}) to clusters of galaxies 
  (triangles; \cite{reiprich}).  The dashed line is the fit to the galaxy 
  (circles only) data; the dotted line is what we expect 
  in $\Lambda$CDM if all the baryons in each halo are identified.}
  \label{fig:BTF}
\end{figure} 

Summing up the mass in stars and gas for a large sample of
spiral galaxies, we obtain the Baryonic Tully-Fisher relation:
\begin{equation}
M_b = 50 V_c^4
\label{eqn:btf}
\end{equation}
with $M_b$ in solar masses and $V_c$ in 
$\textrm{km}\,\textrm{s}^{-1}$ (Fig.~\ref{fig:BTF}).
For an extensive discussion of the effects of various choices
for the stellar mass-to-light ratio on the normalization, slope, and
scatter in this relation, see \cite{btfgood}.  Equation~\ref{eqn:btf}
does well when extrapolated to predict the locations of systems
with $V_c < 50\;\textrm{km}\,\textrm{s}^{-1}$, including
both rotating gas rich dwarf Irr systems (the squares in
Fig.~\ref{fig:BTF}) and pressure supported,
star dominated dwarf spheroidals (Mateo \& Walker; Geha,
private communications).

Extrapolation of the Baryonic Tully-Fisher relation to the cluster
scale does less well, over-predicting the median cluster mass by
a factor of $\sim 2.5$.  Perhaps these are simply different systems,
and such a vast extrapolation is unwarranted.  On the other hand,
coming this close with an extrapolation over several decades
in mass is fairly remarkable.  

That clusters fall a little below the extrapolation of the Baryonic 
Tully-Fisher relation may suggest that our census of baryons
is not complete in these systems.  If so, clusters might be a good 
place to look for some of the missing baryons.  Clusters are not a
large contributor to the total integrated baryonic mass of the universe,
so even if more baryons are lurking there, they are not likely to solve
the missing baryon problem.  Placing them on the extrapolation of
equation~\ref{eqn:btf} would require 
(\textit{very} roughly) 5\% of $\Omega_b$.
While not a large fraction of the total, this is comparable to all the stars
in field galaxies.  Most of the missing baryons could be in the WHIM 
and still leave room for a number of other distinct baryonic reservoirs
like this.

\begin{figure}
 \includegraphics[width=5in]{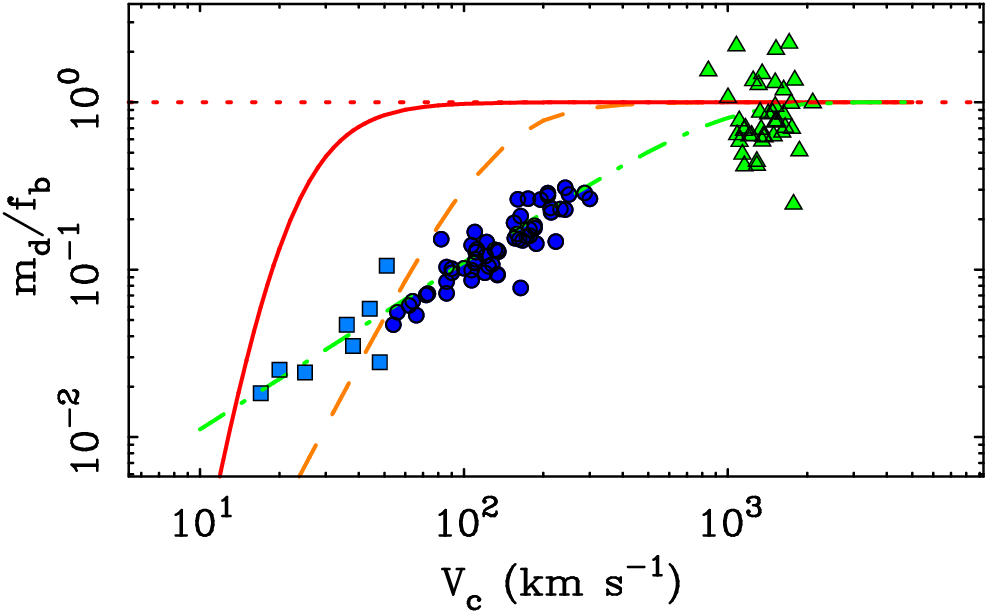}
  \caption{The baryon content as a function of circular velocity.
  Points are galaxy and cluster data, as per Fig.~\ref{fig:BTF}.
  The solid line is the reionization model from \cite{crain} (see also
  \cite{MM04}); the dashed line is the preheating model of
  \cite[Mo \& Mao (2004)]{MoMao}; the dash-dot line is a fit to 
  the data (see text).}
  \label{fig:fb}
\end{figure} 

The CDM mass--velocity relation comes much closer
to the cluster data than does the empirical Baryonic Tully-Fisher
relation.  The difference between it and the median cluster
value is the difference between $f_b = 0.17$ and 0.12.  This near
agreement deteriorates for smaller systems.  The slope of the halo 
mass--virial velocity relation is shallower than the observed baryonic
mass--circular velocity relation.  Even bearing in mind the caveat that 
the halo virial and observed circular velocities may not be simply related,
it is difficult to avoid the conclusion that the observed number of baryons
is systematically less than the expected number as one goes down the
mass function.

If we take the halo mass--virial velocity relation as sacrosanct, then 
it is impossible to avoid the conclusion that $m_d$ varies with $V_c$
(Fig.~\ref{fig:fb}).  Empirically, this variation is well fit by
\begin{equation}
m_d = f_b \tanh\left(\frac{V_c}{V_{\dagger}}\right).
\label{eqn:md}
\end{equation}
Here $V_{\dagger}$ is a scale where $m_d \rightarrow f_b$;
the dash-dot line shown in Fig.~\ref{fig:fb} uses equation~\ref{eqn:md}
with $V_{\dagger} = 900\;\textrm{km}\,\textrm{s}^{-1}$.
This is the scale where the dashed and dotted lines cross in 
Fig.~\ref{fig:BTF} which is where the data transition from looking like 
$\Lambda$CDM on larger scales to looking like MOND on
smaller scales.  Note that something odd is happening on intermediate
scales, as groups in the Local Volume imply $\Omega_m \approx 0.04$
(\cite{K05}).  (Yes, that is $\Omega_m$, not $\Omega_b$.)

In order to derive equation~\ref{eqn:md} or plot Fig.~\ref{fig:fb}, 
it is necessary to make some assumption in order to relate total halo
mass to observed baryonic mass.  For simplicity, I assume 
$V_{vir} = V_c$ here.  While I do not expect this to hold in detail,
plausible adjustments (due, for example, to adiabatic compression) 
will only tweak the details and not alter the basic result.  That 
equation~\ref{eqn:md} returns a value consistent with that for the
Milky Way (equation~\ref{eqn:MW}) suggests that the simple 
assumption is not terrible.

The variation of $m_d$ with $V_c$ implies that there should 
be missing baryons associated with each and every galaxy halo 
(see also Zackrisson, these proceedings).  Equation~\ref{eqn:md} 
is merely a fit to the data.  What should we expect?  Recent work 
gives some idea:  \cite[Mayer \& Moore (2004)]{MM04} 
and \cite{crain} consider
the effects of reionization while \cite[Mo \& Mao (2004)]{MoMao} 
consider the possibility of preheating in analogy with clusters.  
Unfortunately, neither of these types of models matches the observed 
trend (Fig.~\ref{fig:fb}).  In particular, there is no evidence in the data 
for the sudden transition expected around 
$\sim 20\;\textrm{km}\,\textrm{s}^{-1}$ from reionization.  

In the context of $\Lambda$CDM, this merely means that there is
more physics than reionization mediating between the primordial
mass--velocity relation and the observed one.  
Feedback is frequently invoked in this
context, but this word has come to mean many different things.
It remains to be demonstrated whether some flavor of feedback can
be a solution to all these problems.  

There are, in principle, many potential sources of scatter that
should manifest themselves in the Tully-Fisher relation (\cite{MdB98a}),
yet the observed relation has remarkably little scatter.  It is hard to 
imagine how $m_d$ can vary so much between
bright and faint galaxies with virtually zero scatter along the way.
Somehow, a $V_c \approx 50\;\textrm{km}\,\textrm{s}^{-1}$
galaxy must `know' it should have $m_d \approx 0.01$ and never 
0.04, while for a galaxy with 
$V_c \approx 220\;\textrm{km}\,\textrm{s}^{-1}$
the situation is reversed.  The small observed scatter
only happens naturally if $m_d = f_b$ (\cite{btfold}).  
This implies that
essentially all the baryons are accounted for on galaxy scales,
making individual galaxies an unlikely place to look for the
missing baryons.  However, this contradicts a basic tenet of CDM,
as the intrinsic mass--velocity relation differs from 
equation~\ref{eqn:CDM} if $m_d = f_b$.

\section{Conclusions}

I provide two sets of interpretations, one in the context of 
$\Lambda$CDM, the other from an empirical perspective.

\subsection{Conclusions in the context of $\Lambda$CDM}

\begin{itemize}

\item There is no signature in the data due to reionization 
in the vicinity of $\sim 20\;\textrm{km}\,\textrm{s}^{-1}$.  
Low mass galaxies smoothly follow the
extrapolation of the Baryonic Tully-Fisher relation fit to
galaxies with $V_c > 50\;\textrm{km}\,\textrm{s}^{-1}$.

\item The baryon content of halos scales with circular velocity
as summarized by equation~\ref{eqn:md}.

\item Models invoking reionization or pre-heating fail to match the
observed trend in $m_d$.

\item Physics beyond reionization is required to explain the observations.  

\end{itemize}

In order to reconcile the data with $\Lambda$CDM, some physical effect
\textit{must} make the observed baryon fraction $m_d$ the observed
function of circular velocity.  Whatever mechanism is invoked must
operate with remarkably small scatter, posing a fine-tuning problem.
The physics of galaxy formation --- the bridge connecting the primordial
halos found in numerical simulations with real observed galaxies --- 
remains as sound and secure as the Bridge of Death over the
Gorge of Eternal Peril in Monty Python's \textit{Holy Grail}.

\subsection{Empirical conclusions}

\begin{itemize}

\item The observed mass--velocity relation is steeper than predicted
by CDM.

\item The small scatter in the Baryonic Tully-Fisher relation implies
that substantially all the baryonic mass has been accounted for in individual
galaxies.

\item The two preceding items contradict a basic tenet of CDM.

\item Extrapolation of the Baryonic Tully-Fisher relation to cluster scales
suggests that the inventory of baryons in clusters may be incomplete.

\end{itemize}

The last point implies that perhaps some of the missing baryons are hiding 
in clusters in some yet to be identified form.  These need not be a large
fraction of the missing baryons, maybe 5\%.  There is precedence for such 
a situation.  Nowadays it is widely accepted that the hot X-ray emitting 
intra-cluster gas is the dominant baryonic component of rich clusters.
A couple of decades ago, stars were widely thought to be the dominant 
baryonic component.  Though I am at a loss to suggest a plausible state
in which baryons might hide (but see \cite{BJL}), 
this history reminds us that we should not be over-confident about 
the non-existence of matter we haven't yet detected.

 
\begin{acknowledgments}
This material is based upon work supported by the National Science Foundation under Grant No.\ 0505956.  
I would like to thank the organizers, Mike and Jon in particular,
for putting together this delightful conference.
\end{acknowledgments}

\end{document}